\documentclass[twocolumn,showpacs,preprintnumbers,amsmath,amssymb]{revtex4}
\usepackage{dcolumn}% Align table columns on decimal point
\usepackage{bm}% bold math
\usepackage{latexsym}
\usepackage{amsfonts}
\usepackage[dvips]{graphicx}
\usepackage[usenames]{color}
\usepackage{makeidx}
\newcommand{\pa}{\partial}

\newcommand{\ket}{\rangle }
\newcommand{\bra}{\langle }

\newcommand{\ve}{\varepsilon}

\newcommand{\up}{\uparrow}
\newcommand{\dw}{\downarrow}

\begin{document}

%\preprint{APS/123-QED}
\title{
Dielectric Breakdown in a Mott Insulator:
Many-body Schwinger-Landau-Zener Mechanism
studied with a Generalized Bethe Ansatz
}
\author{Takashi Oka and Hideo Aoki}
\address{Department of Physics, University of Tokyo, Hongo, Tokyo 113-0033, 
Japan}

\date{\today}
\begin{abstract}
\noindent 
The nonadiabatic quantum tunneling picture, which may be called 
the many-body Schwinger-Landau-Zener mechanism,
for the dielectric breakdown of Mott insulators 
in strong electric fields is studied in the one-dimensional
Hubbard model.
The tunneling probability is calculated by a metod due to 
Dykhne-Davis-Pechukas with 
an analytical continuation of the 
Bethe-ansatz solution for excited states 
to a non-Hermitian case.
A remarkable agreement with the time-dependent density matrix 
renormalization group result is obtained. 
\end{abstract}

\pacs{05.30.-d, 03.65.Xp, 71.27.+a}
\maketitle

%%%%%%%%%%%%%%%%%%%%%%%%%%%%%%%%%%%%%%%%%%%%%%%%%%%%%%%%%%%%%%%%%%%
%{\it Introduction ---} 
%%%%%%%%%%%%%%%%%%%%%%%%%%%%%%%%%%%%%%%%%%%%%%%%%%%%%%%%%%%%%%%%%%%
Among nonequilibrium and nonlinear transport 
phenomena in correlated electron systems, 
dielectric breakdown (destruction 
of insulating states due to strong electric fields) 
is one of the most basic. 
In Mott insulators, electrons freeze their 
motion due to strong repulsive interaction \cite{Imada1998}, 
and in equilibrium an introduction of carriers in a Mott insulator 
leads to interesting quantum states such as
high Tc superconductivity in 2D 
or Tomonaga-Luttinger liquids in 1D. 
Now, it is an intriguing problem to ask
how nonequilibrium carriers behave when 
electrons in a Mott insulator
start to move in strong enough electric fields.

The nonequilibrium phase transition 
from Mott insulators to metals by electric fields
has been studied
in the condensed-matter physics\cite{tag,Oka2003,Oka2005a,Fukui}. 
More recently, the problem is attracting interest in the cold atom physics, 
where novel realization of the Mott insulator
has been achieved in 
bosonic\cite{Greiner2002,PhysRevLett.91.230406,PhysRevLett.93.140406} as well as in fermionic\cite{U.Schneider12052008} systems.
The many-body Landau-Zener mechanism for 
dielectric breakdown has been proposed 
for fermionic systems in ref.~\cite{Oka2003}, and 
for bosonic systems in ref.~\cite{witthaut:063609}. 
The correspondence between the Landau-Zener mechanism
and the Schwinger mechanism\cite{Schwinger1951} in strong-field QED
as well as the relation between the Heisenberg-Euler effective Lagrangian
and the nonadiabatic geometric phase was 
given in ref.~\cite{Oka2005a} (see also ref.~\cite{Green2005}).
In ref.~\cite{Oka2005a} an extensive numerical calculation was performed 
to obtain the electric-field induced nonequilibrium 
phase diagram. 
One important prediction of the 
Schwinger-Landau-Zener picture is that the 
threshold electric field $E_{\rm th}$ for the breakdown 
is related to the charge gap 
$\Delta(U)$ as $eE_{\rm th}\propto \Delta^2(U)$, 
which is much smaller than a 
naive guess of $eE_{\rm th}\sim  U$, i.e., 
the energy offset between 
neighboring sites in a tilted potential 
($U$: the onsite repulsion). 
Such lowering of the threshold was 
experimentally observed by 
Taguchi {\it et al.}\cite{tag} 
who measured the $I$-$E$ characteristics in
a one-dimensional Mott insulator, where 
a quantum origin of the
breakdown was suggested from a threshold that remains 
finite in the zero-temperature limit.  
In cold atoms, the effect of the potential gradient was studied~\cite{Greiner2002} to 
probe the excitation spectrum
(they use the relation $eE_{\rm th}\sim U$ to interpret their results).

However, the Schwinger-Landau-Zener theories have a snag in
many-body systems: 
As explained below eqn.~(\ref{eq:Flz}), 
the Landau-Zener threshold contains a factor that depends on the 
system size and diverges in the 
thermodynamic limit, i.e., no breakdown would take place
in bulk systems, which contradicts with intuition.
The purpose of the present paper is to resolve this puzzle, 
where an analytic expression for the threshold field strength
valid in the thermodynamic limit is presented.  
This has been achieved 
by deriving the quantum transition probability utilizing a 
method due to 
Dykhne-Davis-Pechukas (DDP) formalism
which enables us to treat quantum tunneling beyond the 
Landau-Zener picture\cite{Dykhne1962,DavisPechukas1976}. 

The present approach has another virtue:  
Besides the quantum tunneling approach, 
there is a non-Hermitian approach 
studied by Fukui and Kawakami\cite{Fukui}, 
where the authors incorporated 
phenomenologically the effect of  
electric fields as differing left and right 
hopping terms (for non-Hermitian models 
see also \cite{PhysRevLett.77.570,NakamuraHatano06}). 
However, the relation to experiments was not too clear, since 
a direct connection between the ratio of the left- and right-going 
hoppings with the applied field strength was not given. 
In the present derivation, the non-Hermitian formalism 
emerges naturally, and the two apparently unrelated theories 
(i.e., Schwinger-Landau-Zener and non-Hermitian) 
are shown to be in fact intimately related.  
Indeed, the transition probability  in the DDP 
is calculated with an analytic continuation of the 
solution of the time-dependent Hamiltonian onto 
a complex time, and the Hubbard model in an electric field is 
mapped onto a non-Hermitian model.  
In order to complete the calculation,
we need the information on excited states.  
This has been achieved here for the 1D Hubbard model 
with a {\it non-Hermitian generalization of the 
Bethe-ansatz}\cite{Lieb:1968AM,Hubbardbook} excited states, i.e.,
the string 
solutions\cite{PhysRevB.9.2150,Ovchinnikov70,Takahashi72,Woynarovich1982}. 
The present result turns out to agree with the time-dependent 
density matrix renormalization group result\cite{Oka2005a} with a 
remarkable accuracy.  

%%%%%%%%%%%%%%%%%%%%%%%%%%%%%%%%%%%%%%%%%%%%%
\begin{figure}[ht]
\centering 
\includegraphics[width=8cm]{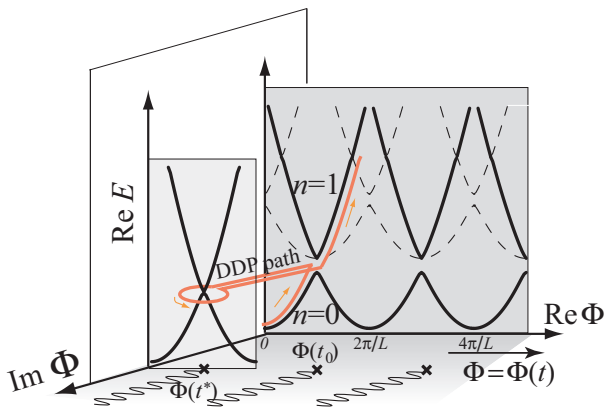}
\caption{(color online)
Many-body energy levels against the complex AB flux $\Phi$ 
for a finite, half-filled 1D Hubbard model 
($L=10,\;N_\up=N_\dw=5$, $U=0.5$).  
Only charge excitations are plotted. 
Quantum tunneling occurs between the groundstate (labeled as $n=0$) 
and a low-lying excited state ($n=1$)
as the flux $\Phi(t)=Ft$ increases 
on the real axis, while 
the tunneling is absent for the 
states plotted as dashed lines.  
The wavy lines starting from the singular points 
($\times$) at $\Phi(t^*)$ 
represent the branch cuts for different Riemann surfaces, 
along which the solutions $n=0$ and $n=1$ are connected. 
In the DDP approach, the tunneling factor is 
calculated from the dynamical phase associated
with adiabatic time evolution (DDP path) 
that encircles a gap-closing  point at
$\Phi(t^*)$ on the complex $\Phi$ plane. 
}
\label{fig1}
\end{figure}
%%%%%%%%%%%%%%%%%%%%%%%%%%%%%%%%%%%%%%%%%%%%

%%%%%%%%%%%%%%%%%%%%%%%%%%%%%%%%%%%%%%%%%%%%%%%%%%%%%%%%%%%%%%%%%%%
%{\it Quantum tunneling and non-Hermitian Hubbard model ---} 
%%%%%%%%%%%%%%%%%%%%%%%%%%%%%%%%%%%%%%%%%%%%%%%%%%%%%%%%%%%%%%%%%%%
Here we consider the time evolution of 
electrons in a strong electric field
$E$ for the one-dimensional Hubbard model,
\begin{equation}
H = -\sum_{i,\sigma}
\left( e^{i\Phi(t)}c_{i+1\sigma}^\dagger c_{i\sigma}
+\mbox{h.c.} \right)+U\sum_{i}n_{i\up}n_{i\dw},
\label{eq:Hamiltonian}
\end{equation}
where the electric field is introduced by a 
time-dependent phase $\Phi(t)=Ft$ 
with $F=eE$ switched on at $t=0$. 
This is one obvious way of introducing an electric field 
through Faraday's law.  
We have taken the absolute value of the hopping as the 
unit of energy. 
We study a half-filled, nonmagnetic case with 
numbers of electrons $N_\up=N_\dw=L/2$ with $L$ the 
total number of sites.
The Mott-insulator groundstate becomes unstable
when the electric field
becomes strong enough, for which charge 
excitations take place 
due to nonadiabatic quantum tunneling \cite{Oka2003}. 
In order to describe the process we introduce the adiabatic levels 
$|\psi_n(\Phi)\ket$
that satisfy $
H(\Phi)|\psi_n(\Phi)\ket=E_n(\Phi)|\psi_n(\Phi)\ket
$
with $n=0,1,\ldots$, where $n=0$ corresponds to the groundstate. 
We neglect spin excitations to concentrate on 
charge excitations. 
The time evolution for $t>0$ is described by
the time dependent Schr\"odinger 
equation, $i\frac{d}{dt}|\psi(t)\ket=
H(t)|\psi(t)\ket$, with initial state 
$|\psi(t=0)\ket=|\psi_0(\Phi=0)\ket$.
Figure~\ref{fig1} plots the adiabatic energy levels 
obtained by exact diagonalization for a small system.
Nonadiabatic quantum tunneling between the groundstate and the lowest charge-excited state is most relevant 
(while the transition to the state represented by dashed 
lines is absent due to symmetry reasons). 
The adiabatic levels are periodic in $\Phi$ 
with a period $2\pi/L$, so that the tunneling from
the groundstate to the excited state 
repeatedly occurs with a time interval $T=2\pi /FL$.
We define the tunneling factor between 
the two states by $\gamma_{0\to 1}$
which is related to the transition probability
$P=e^{-\gamma_{0\to 1}}$ for a 
single tunneling event.  
The solution of the 
time-dependent Schr\"odinger equation behaves as
$|\psi(mT)\ket\sim (1-e^{-\gamma_{0\to 1}})^{m/2}e^{i\alpha(t)}
|\psi_0(t)\ket$ with a phase factor $\alpha$, 
and the groundstate decay rate $\Gamma$ defined by
$|\bra\psi_0(\Phi(t))|\psi(t)\ket|^2=e^{-\Gamma t}$
becomes\cite{Oka2005a}
%\begin{equation}
$\Gamma/L\sim -\frac{F}{2\pi}\ln(1-e^{-\gamma_{0\to 1}})$.
%\end{equation}
A naive estimate for the 
tunneling factor can be made by approximating
the Hamiltonian in the vicinity of the transition 
by a Landau-Zener form, 
$H^{\rm LZ}=(^{vt\;\Delta/2}_{\Delta/2\;-vt})$,
which leads to a threshold behavior 
with threshold $F_{\rm th}^{\rm LZ}$ given by
\cite{Oka2003,Oka2005a}
\begin{eqnarray}
\gamma^{{\rm LZ}}_{0\to 1}=\pi\frac{F_{\rm th}^{\rm LZ}}{F},\;\;\;
F_{\rm th}^{\rm LZ}=\frac{(\Delta/2)^2}{v},
\label{eq:Flz}
\end{eqnarray}
where $\Delta$ is the charge gap (Mott gap) \cite{Lieb:1968AM}, 
and $v$ is the slope of the adiabatic levels 
($v\sim 2$ when $U$ is small and the system size is small).   
However, this expression should fail 
when the system size exceeds
the localization length\cite{Stafford1993}, 
since the slope vanish $v\to 0$ and the levels become flat against $\Phi$.
Then, the transition probability also vanishes. 
But this obviously contradicts with a physical intuition that 
dielectric breakdown should take place 
in infinite systems.  
The point is that quantum tunneling
can take place even when the levels are 
flat \cite{RozenZener}.

In order to resolve this problem we introduce the DDP method 
which accommodates the thermodynamic limit 
as we shall see. 
In the general formalism of DDP 
the solution of the Schr\"odinger equation 
is extended to complex time;  
The tunneling process is described by an adiabatic
evolution of the wave function along a path 
in the complex plane (DDP path in Fig.~\ref{fig1},
displayed for a finite system for clarity).
The DDP path encircles the point
$t^*$ (exceptional point) on the
complex $t$ plane at which the two energy levels 
cross, i.e., $E_1(\Phi(t^*))=E_0(\Phi(t^*))$.
There is a branch cut starting from $t^*$ at which the
two Riemann surfaces corresponding to $E_0$ and $E_1$ merge, 
and along a path encircling $t^*$ the solution $|\psi_0\ket$
is deformed into the excited state $\propto|\psi_1\ket$ with a 
proportionality factor determined by the complex dynamical phase.
This gives a DDP tunneling probability $P=e^{-\gamma^{{\rm DDP}}_{0\to 1}}$
with 
\cite{Dykhne1962,DavisPechukas1976,PhysRevA.59.4580,Wilkinson2000}
\begin{eqnarray}
\gamma^{{\rm DDP}}_{0\to 1}=2\mbox{Im}S_{0,1}/\hbar,
\label{eq:DDP}
\end{eqnarray}
where $S$ is the dynamical phase given by
\begin{eqnarray}
S_{0,1}=\int_{t_0}^{t^*}dt'[E_1(\Phi(t'))-E_0(\Phi(t'))],
\label{eq:DDPdiff}
\end{eqnarray}
with $t_0$ the starting point on the real axis.

\begin{figure}[b]
\centering 
\includegraphics[width=8.5cm]{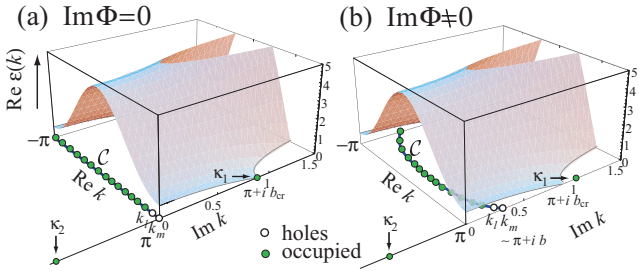}
\caption{(color online)
Schematic configurations 
(displayed here for $U=4.0$) of the
charge rapidities for 
the lowest charge-excited state $|\psi_1(i\Psi)\ket$ 
for $\Psi=0$(a), and for a finite $\Psi$(b).  
$\mathcal{C}$ corresponds to the 
groundstate continuum with occupied states (green circles, printed grey). 
In the excited state, two holes $k_l,\; k_m$ (open circles) 
near $\pi+ib$
appear in the continuum, 
while two rapidities $\kappa_1,\;\kappa_2$ outside the continuum $\mathcal{C}$ 
are occupied. 
The surface represents the real part of the excitation energy $\mbox{Re} \, \ve(k)$ 
(plotted here for $\mbox{Re} \, \ve(k)\ge 0$, $\mbox{Im} \, k>0$) 
which gives the energies of holes at $k_l, k_m$. 
At $\Psi=0$, $\kappa_1, \kappa_2$ sit on the $\mbox{Re} \, \ve(k)=0$ curve. 
}
\label{rapidity}
\end{figure}

We want to apply the DDP method (eqns.~(\ref{eq:DDP}),~(\ref{eq:DDPdiff})) 
to the Hubbard model, which 
means that we have to 
analytically continue the 
solutions to complex $\Phi$ for the first excited state
($E_1$) as well as 
for the groundstate ($E_0$). 
The Hubbard model with the phase factor 
(eqn.~(\ref{eq:Hamiltonian})) is exactly solvable 
with the Bethe ansatz method
(see for example \cite{kusakabe}). 
This remains the case,
for the groundstate, 
even when $\Phi$ is complex 
as demonstrated by Fukui and Kawakami \cite{Fukui}. 
However, we have to extend the procedure to the 
excited states (Fig.~\ref{rapidity}), which is feasible with 
Woynarovich's method \cite{Woynarovich1982},
where our goal is to 
calculate the energy difference $E_1(\Phi)-E_0(\Phi)$
for complex $\Phi$ and perform the 
integral along the DDP path.
The DDP path (Fig.\ref{fig1}) for the Hubbard model 
starts from $\Phi_0=\pi/L$ 
and ends at 
$\Phi_*=\pi/L+i\Psi_{\rm cr}$, where 
$\Psi_{\rm cr}$ is the value at which the 
gap closes \cite{Fukui}.  In the 
large $L$ limit the path lies on the imaginary axis.

%%%%%%%%%%%%%%%%%%%%%%%%%%%%%%%%%%%%%%%%%%%%%
%%%%%%%%%%%%%%%%%%%%%%%%%%%%%%%%%%%%%%%%%%%%

We start with the Lieb-Wu Bethe-ansatz equation for an $L$-site Hubbard model with an imaginary $\Phi=i\Psi$,
\begin{eqnarray}
&&Lk_j = 2\pi I_j+iL\Psi-\sum_{\alpha=1}^{N_\dw}\theta(\sin k_j-\lambda_\alpha),
\label{eq:LW1}\\
&&
\sum_{j=1}^{L}\theta(\sin k_j-\lambda_\alpha)
=2\pi J_\alpha-\sum_{\beta=1}^{N_\dw}
\theta\left(\frac{\lambda_\alpha-\lambda_\beta}{2}\right)
\label{eq:LW2},
\end{eqnarray}
where $k_j \;(\lambda_\alpha)$ are the 
charge (spin) rapidities, 
$\theta(x)=-2\mbox{arctan}(x/u)$ with $u=U/(4t)$ 
is the two-body phase shift, and 
$I_j\equiv N_\dw/2\;(\mbox{mod}~1),
\;J_\alpha\equiv (N-N_\dw+1)/2\;(\mbox{mod}~1)$.

In the infinite-size limit, the Lieb-Wu equation
for a finite $\Psi$
can be solved with the analytically continued
charge and spin distribution functions \cite{Fukui}.  
If we introduce the counting functions $z_c(k_j)=I_j/L$ and 
$z_s(\lambda_\alpha)=J_\alpha/L$, 
the Lieb-Wu equation in the bulk limit reads
\begin{eqnarray}
z_c(k)&=&\frac{k}{2\pi}-\frac{i\Psi}{2\pi}-\frac{1}{2\pi}
\int_{\mathcal{S}}d\lambda \, \theta(\sin k-\lambda)\sigma^*(\lambda),
\label{eq:LiebWus2}\\
z_s(\lambda)&=&\frac{1}{2\pi}\int_{\mathcal{C}}dk \, \theta(\sin k-\lambda)
\rho^*(k)
\nonumber \\ 
&&+\frac{1}{2\pi}\int_{\mathcal{S}}d\lambda' \, \theta
\left(\frac{\lambda-\lambda'}{2}\right)\sigma(\lambda'),
\label{eq:LiebWuc2}
\end{eqnarray}
where the distribution functions are defined by
$
\rho(k) = \pa_kz_c(k)
,\;
\sigma(\lambda) = \pa_\lambda z_s(\lambda)$, 
and $\sigma^*,\sigma, \rho^*$ are explained around 
eqn.~(\ref{eq:rhodef}) below. 
%and replacing them with $\sigma_0$ and $\rho_0$ 
%gives the Lieb-Wu equation for the groundstate. 
The contours $\mathcal{C}$ and $\mathcal{S}$,
i.e., the continuum limit of the charge ($\mathcal{C}$)
and spin ($\mathcal{S}$) 
rapidities' positions, are of great importance. 
In fact, for the groundstate, the paths are determined such that the 
conventional solution\cite{Lieb:1968AM}, 
$\rho_0(k)=\frac{1}{2\pi}+\frac{1}{2\pi}
\cos k\int_0^\infty\frac{e^{-u\omega}}{\cosh u\omega}
J_0(\omega)\cos(\omega\sin k)d\omega$
 and $\sigma_0(\lambda)=\frac{1}{2\pi}\int_0^\infty
\frac{J_0(\omega)\cos\omega\lambda}
{\cosh u\omega}d\omega$ with $J_n$ Bessel's function, extended to 
complex $k$ and $\lambda$ solves 
eqns.~(\ref{eq:LiebWus2}),~(\ref{eq:LiebWuc2}). 
This determines the end point
of contour $\mathcal{C}$, which we 
denote $\pm \pi+ib$ (Fig. \ref{rapidity}), where 
$b$ is an increasing function of $\Psi$ 
satisfying \cite{Fukui}
\begin{equation}
\Psi = b-i\int_{-\infty}^\infty d\lambda
\theta(\lambda+i\sinh b)\sigma_0(\lambda).
\label{Psivsb}
\end{equation}
We denote the end point corresponding to 
$\Psi=\Psi_{\rm cr}$ to be $b=b_{\rm cr}$. 
The end point of $\mathcal{S}$ is $\lambda=\pm \pi$.

\begin{figure}[t]
\centering 
\includegraphics[width=8.5cm]{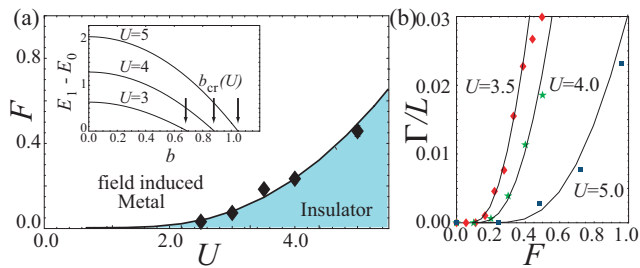}
\caption{(a) The threshold field strength $F$ against $U$ 
obtained by the present DDP formalism 
(solid line; eqn.~(\ref{eq:Fthbethe})).  
Inset: The energy difference between the groundstate and 
the excited state against $b$ for various values of $U$.
(b) The decay rate $\Gamma$ of the groundstate against the electric field $F$ obtained by the DDP formalism (solid line; eqn.(\ref{eq:gamma})).  
In (a) and (b), the symbols represent the 
time-dependent DMRG result \cite{Oka2005a}.
}
\label{threshold}
\end{figure}

Woynarovich's construction \cite{Woynarovich1982}  
of charge excitations can be
applied to the non-Hermitian case ($\Psi \neq 0$) 
with the same contours 
$\mathcal{C}$, $\mathcal{S}$ as in the groundstate.
The idea is to remove two charge rapidities $k_l,\;k_m$ 
from $\mathcal{C}$ 
and one spin rapidity $\lambda_{N_\dw/2}$ 
from $\mathcal{S}$ to 
place them on the complex $k$ and $\lambda$ planes 
at positions $\kappa_1,\;\kappa_2$ and 
$\Lambda$, respectively (Fig.~\ref{rapidity}) 
in such a way that the Lieb-Wu
equation is satisfied, which yields
\begin{eqnarray}
\sin(\kappa_{1,2})=\Lambda\pm iu, \quad \Lambda=(\sin k_l+\sin k_m)/2.
\label{eq:kappa}
\end{eqnarray}
With these parameters, the Lieb-Wu equation (\ref{eq:LiebWuc2})
for charge excitations
can be solved by 
\begin{eqnarray}
\sigma(\lambda) &=& \sigma_0(\lambda)-\frac{1}{LU}\left\{ 
\frac{1}{\cosh[(\lambda-\sin k_l)\pi/2u]}\right. \nonumber \\
&+&\left. \frac{1}{\cosh[(\lambda-\sin k_m)\pi/2U]}
 \right\},
\label{eq:rhodef}\\
\rho(k) &=& \rho_0(k)+\frac{1}{2\pi L}\cos k
\frac{u}{u^2+(\sin k-\Lambda)^2}\nonumber \\&&
-\frac{\cos k}{2\pi L}
\int_0^\infty\frac{e^{-\omega u}}{\cosh \omega u}
\{\cos[\omega (\sin k-\sin k_l)]
\nonumber \\&&
+\cos[\omega (\sin k-\sin k_m)]\}d\omega,
\end{eqnarray}
with which we can define 
$\sigma^*(\lambda)=\sigma(\lambda)+(1/L)\delta(\lambda-\Lambda),\;
\rho^*(k)=\rho(k)-(1/L)\delta(k-k_l)-(1/L)\delta(k-k_m)$
appearing above.
We note that these equations are 
identical with Woynarovich's, which is 
natural since the operations 
$\pa_k, \pa_\lambda$ do not pick up $\Psi$, while 
$\Psi$ controls the integration path via eq.(\ref{Psivsb}).
The energy of the excited state
can be calculated from $\rho^*(k)$,
which gives
%\begin{eqnarray}
$E_1(\Psi)-E_0(\Psi)=\ve(k_l)+\ve(k_m)$
%\label{eq:energydiff}
%\end{eqnarray}
with $E_0$ the groundstate energy, and 
the $\ve(k)$ given as
\begin{eqnarray}
\ve(k)&=&2u+2\cos(k)
\label{eq:vek}\\
&&+2\int_0^\infty
\frac{e^{-u\omega}}{\omega\cosh u\omega}
J_1(\omega)\cos(\omega\sin k)d\omega.\nonumber
\end{eqnarray}
The lowest excited state is given by 
setting $k_l,k_m \simeq \pi+ib$ in the above solution (Fig.\ref{rapidity}). 
We can specify the deformation of the Bethe ansatz solution 
along the DDP path (Fig.~\ref{fig1}) as follows.
As $\Psi$ becomes finite, the 
end points of $\mathcal{C}$, i.e., $\pm \pi+ib$,
move along the imaginary axis 
until $b$ reaches $b_{\rm cr}$ 
at which the gap closes, i.e.,
$E_{1}-E_{0}=0$ (Fig.~\ref{threshold}, inset)\cite{Fukui}. 
Meanwhile,  $\mbox{Im} \, \kappa_1$ and $\mbox{Im} \, \kappa_2$
increase with $\Psi$, 
where $\kappa_2$ in particular touches the real axis 
at the critical point. 
From the DDP formula (eqns.~(\ref{eq:DDP}),~(\ref{eq:DDPdiff})),
the quantum tunneling probability
$P=e^{-\pi F_{\rm th}^{\rm DDP}/F}$ 
has the threshold electric field, 
\begin{eqnarray}
&&F_{\rm th}^{\rm DDP}
 = \frac{2}{\pi}\int_0^{b_{\rm cr}}(E_1-E_0)\frac{d\Psi}{db}
db\nonumber\\
&=& 
\frac{2}{\pi}\int_0^{\sinh^{-1}u}
4\left[u-\cosh b+\int_{-\infty}^\infty
d\omega\frac{e^{\omega\sinh b}J_1(\omega)}
{\omega (1+e^{2u|\omega|})}
\right]
\nonumber\\
&&\times\left[
1-\cosh b
\int_0^\infty d\omega\frac{J_0(\omega)\cosh(\omega \sinh b)}{1+
e^{2 u\omega}}\right]db.
\label{eq:Fthbethe}
\end{eqnarray}
%\end{widetext}
Its $U$-dependence is plotted e in Fig.~\ref{threshold}~(a) 
(solid line), which confirms the 
collective nature of the breakdown 
(i.e., the threshold much smaller than a naive 
$U$). In other words, the tunneling 
takes place not between neighboring sites, 
but over an extended region
due to a {\it leakage of the many-body wave function},
where the size is roughly 
the localization length \cite{Stafford1993}. 

Let us now compare the present analytical result with 
the numerical one in Fig.~\ref{threshold}~(a), which 
plots $F_{\rm th}^{\rm DDP}$ 
along with the threshold obtained  by the 
time-dependent density-matrix renormalization group (DMRG)
for an $L=50$, open Hubbard chain \cite{Oka2005a}. 
The agreement between the analytical and numerical 
results is excellent.  

%%%%%%%%%%%%%%%%%%%%%%%%%%%%%%%%%%%%%%%%%%%%%
%%%%%%%%%%%%%%%%%%%%%%%%%%%%%%%%%%%%%%%%%%%%

Finally, let us say a few words about 
the dynamics that takes place after the 
electric field exceeds the threshold. 
There are infinitely many excited states
whose energies are larger but near $|\psi_1\ket$'s, 
and tunneling becomes also activated to these 
states. The net tunneling to such states 
is incorporated in the groundstate decay rate 
$\Gamma/L$ (defined above eqn.(\ref{eq:Flz})). 
This quantity has been numerically calculated with the time-dependent 
DMRG in ref.\cite{Oka2005a}, where 
the single-tunneling formula reduced by 
an empirical factor $a<1$,
\begin{equation}
\Gamma/L=-\frac{aF}{2\pi}\ln[1-\exp(-\pi F_{\rm th}^{\rm DDP}/F)],
\label{eq:gamma}
\end{equation}
is found to describe the numerical result.  
The present DDP result again exhibits
an excellent agreement with the
numerical one (Fig.~\ref{threshold}(b)).
This implies that the tunneling to higher excited states do not
change the threshold, while the decay rate is reduced 
due to the pair-annihilation processes\cite{Oka2005a}. 
We note that the decay rate is an experimental observable
which can be obtained from the delay time of the current
(the production rate in ref.~\cite{tag} Fig.~4), 
and the present theory is consistent with the experimental result.
The nature of  the nonequilibrium steady state above the threshold 
is an interesting problem, which will be addressed elsewhere 
where an electron avalanche effect is evoked  for 
the metallization.

%Note that the agreement 
%between the numerical and DDP thresholds revealed here 
%indicates that tunneling processes to higher excited states
%do not change the threshold significantly, which 
%may be because they are
%activated only after the tunneling to the
%lowest excited state is activated.

%However, there should be a back-reaction;
%As the densities of doubly occupied sites 
%and empty sites become large, 
%pair-annihilation processes are activated as well. 
%This can be viewed as a quantum tunneling 
%process from the excited states
%back to the groundstate \cite{Oka2005a}
%and if the quantum interference becomes
%dominant, a dynamical localization 
%may take place \cite{Oka2004a}.

In conclusion, we have shown that 
the DDP theory of quantum tunneling combined with 
a generalized Bethe ansatz describes 
the nonlinear transport and dielectric breakdown of 
the 1D Mott insulator.  
This is the first analytical result obtained on 
nonequilibrium properties in correlated electron system, 
and the DDP method is expected to have 
potential applicability to many other
models and problems. 
We wish to thank Mitsuhiro Arikawa, Yasuhiro Hatsugai and Takahiro Fukui for fruitful discussions, and Seiji Miyashita for bringing our attention to 
\cite{RozenZener}. 
HA was supported by a Grant-in-Aid for Scientific Research on Priority Area
``Anomalous quantum materials", TO by a Grant-in-Aid for Young Scientists (B) 
from MEXT.

%\bibliographystyle{apsrev.bst}
%\bibliographystyle{unsrt.bst}
%\bibliography{c:/Physics/ref.bib}
%\printindex

\end{document}